\documentstyle[PASJadd]{PASJ95}
\input epsf.sty

\begin{document}
\title{Infrared Imaging of the Gravitational Lens PG~1115+080 with the Subaru Telescope}

\author{
Fumihide {\sc Iwamuro},$^1$
Kentaro {\sc Motohara},$^1$
Toshinori {\sc Maihara},$^1$
Jun'ichi {\sc Iwai},$^1$\\
Hirohisa {\sc Tanabe},$^1$
Tomoyuki {\sc Taguchi},$^1$
Ryuji {\sc Hata},$^1$
Hiroshi {\sc Terada},$^1$
Miwa {\sc Goto},$^1$\\
Shin {\sc Oya},$^5$
Masayuki {\sc Akiyama},$^{2,3}$
Hiroyasu {\sc Ando},$^4$
Tetsuo {\sc Aoki},$^5$
Yoshihiro {\sc Chikada},$^3$\\
Mamoru {\sc Doi},$^6$
Takeo {\sc Fukuda},$^7$
Masaru {\sc Hamabe},$^8$
Masahiko {\sc Hayashi},$^2$
Saeko {\sc Hayashi},$^2$\\
Toshihiro {\sc Horaguchi},$^9$
Shinichi {\sc Ichikawa},$^{10}$
Takashi {\sc Ichikawa},$^{11}$
Masatoshi {\sc Imanishi},$^4$\\
Katsumi {\sc Imi},$^2$
Motoko {\sc Inata},$^4$
Shuzo {\sc Isobe},$^4$
Yoichi {\sc Itoh},$^2$
Masanori {\sc Iye},$^4$
Norio {\sc Kaifu},$^2$\\
Yukiko {\sc Kamata},$^7$
Tomio {\sc Kanzawa},$^2$
Hiroshi {\sc Karoji},$^4$
Nobunari {\sc Kashikawa},$^4$
Taichi {\sc Kato},$^3$\\
Naoto {\sc Kobayashi},$^2$
Yukiyasu {\sc Kobayashi},$^4$
Keiichi {\sc Kodaira},$^{12}$
George {\sc Kosugi},$^2$\\
Tomio {\sc Kurakami},$^2$
Yoshitaka {\sc Mikami},$^4$
Akihiko {\sc Miyashita},$^2$
Takashi {\sc Miyata},$^4$\\
Satoshi {\sc Miyazaki},$^7$
Yoshihiko {\sc Mizumoto},$^4$
Masao {\sc Nakagiri},$^4$
Koich {\sc Nakajima},$^{13}$\\
Kyoko {\sc Nakamura},$^7$
Kyoji {\sc Nariai},$^{12}$
Eiji {\sc Nishihara},$^{10}$
Jun {\sc Nishikawa},$^4$
Shiro {\sc Nishimura},$^{12}$\\
Tetsuo {\sc Nishimura},$^2$
Tetsuo {\sc Nishino},$^7$
Kunio {\sc Noguchi},$^4$\\
Takeshi {\sc Noguchi},$^4$
Jun'ichi {\sc Noumaru},$^2$
Ryusuke {\sc Ogasawara},$^2$
Norio {\sc Okada},$^7$\\
Kiichi {\sc Okita},$^4$
Koji {\sc Omata},$^2$
Norio {\sc Oshima},$^7$
Masashi {\sc Otsubo},$^2$
Goro {\sc Sasaki},$^7$\\
Toshiyuki {\sc Sasaki},$^2$
Maki {\sc Sekiguchi},$^{14}$
Kazuhiro {\sc Sekiguchi},$^2$
Ian {\sc Shelton},$^2$
Chris {\sc Simpson},$^2$\\
Hiroshi {\sc Suto},$^1$
Hideki {\sc Takami},$^4$
Tadafumi {\sc Takata},$^2$
Naruhisa {\sc Takato},$^4$
Motohide {\sc Tamura},$^4$\\
Kyoko {\sc Tanaka},$^4$
Wataru {\sc Tanaka},$^4$
Daigo {\sc Tomono},$^2$
Yasuo {\sc Torii},$^4$
Tomonori {\sc Usuda},$^2$\\
Koichi {\sc Waseda},$^7$
Juni'chi {\sc Watanabe},$^{15}$
Masaru {\sc Watanabe},$^{10}$
Masafumi {\sc Yagi},$^4$\\
Takuya {\sc Yamashita},$^2$
Yasumasa {\sc Yamashita},$^{12}$
Naoki {\sc Yasuda},$^4$
Michitoshi {\sc Yoshida},$^4$\\
Shigeomi {\sc Yoshida},$^{16}$
and
Masami {\sc Yutani},$^4$
\\ [12pt]
$^1${\it Department of Physics, Kyoto University, Kitashirakawa, Kyoto 606-8502}\\
{\it E-mail(FI): iwamuro@cr.scphys.kyoto-u.ac.jp}\\
$^2${\it Subaru Telescope, National Astronomical Observatory, 650 North Aohoku Place, Hilo, HI 96720, USA}\\
$^3${\it Department of Astronomy, Kyoto University, Kyoto 606-8502}\\
$^4${\it Optical and Infrared Astronomy Division, National Astronomical Observatory, Mitaka, Tokyo 181-8588}\\
$^5${\it Communications Research Laboratory, Koganei, Tokyo 184-8975}\\
$^6${\it Department of Astronomy, The University of Tokyo, Bunkyo-ku, Tokyo 113-0033}\\
$^7${\it Advanced Technology Center, National Astronomical Observatory, Mitaka, Tokyo 181-8588}\\
$^8${\it Institute of Astronomy, The University of Tokyo, Mitaka, Tokyo 181-8588}\\
$^9${\it National Science Museum, Taito-ku, Tokyo 110-8718}\\
$^{10}${\it Astronomical Data Analysis Center, National Astronomical Observatory, Mitaka, Tokyo 181-8588}\\
$^{11}${\it Astronomical Institute, Tohoku University, Aoba-ku, Sendai 980-8578}\\
$^{12}${\it National Astronomical Observatory, Mitaka, Tokyo 181-8588}\\
$^{13}${\it Hitotsubashi University, Kunitachi, Tokyo 186-8601}\\
$^{14}${\it Cosmic Ray Research Laboratory, The University of Tokyo, Tanashi, Tokyo 188-8502}\\
$^{15}${\it Public Relation Office, National Astronomical Observatory, Mitaka, Tokyo 181-8588}\\
$^{16}${\it Kiso Observatory, The University of Tokyo, Mitake, Nagano 397-0101}\\
}
\abst{
We present high spatial resolution images of the gravitational-lens 
system PG~1115+080 taken with the near-infrared camera (CISCO) on the 
Subaru telescope. The FWHM of the combined image is $0.\hspace{-2pt}''32$ in the 
$K'$-band, yielding spatial resolution of $0.\hspace{-2pt}''14$ after a 
deconvolution procedure. This is a first detection of an extended 
emission adjacent to the A1/A2 components, indicating the presence of 
a fairly bright emission region with a characteristic angular radius 
of $\sim$ 5~mas (40~pc). The near-infrared image of the Einstein ring 
was extracted in both the $J$ and $K'$ bands. The $J-K'$ color is found 
to be significantly redder than that of a synthetic model galaxy 
with an age of 3~Gyr, the age of the universe at the quasar redshift.}

\kword{galaxies: structure --- gravitational lensing --- 
quasars: individual (PG 1115+080)}

\maketitle
\thispagestyle{headings}

\clearpage
\section
{Introduction}

The quadruply imaged $z$ = 1.722 quasar PG~1115+080 (Weymann et al.\ 1980) is one 
of the gravitational lens systems in which time delays have been 
measured (Schechter et al.\ 1997; Barkana 1997). Astrometric and photometric 
observations have been made by HST (Kristian et al.\ 1993; Impey et al.\ 1998) and by 
ground-based telescopes (Courbin et al.\ 1997). As a result, they have 
precisely determined the radial offset of each lensed component from the lensing 
galaxy, which is located at a redshift of 0.311 (Kundic et al.\ 1997; Tonry 1998). 
Restrictions on the Hubble constant were obtained on the basis 
of these observational materials along with appropriate lens models 
(Courbin et al.\ 1997; Kundic et al.\ 1997; Keeton, Kochanek 1997; 
Saha, Williams 1997; Impey et al.\ 1998). 

A gravitational-lens system also offers an opportunity to study the 
detailed structure of the lensed object, because the lens effect 
serves as a {\it natural telescope} which magnifies and brightens the image 
of the lensed object, although the lensed images are distorted. 

Recently, the infrared Einstein ring of PG~1115+080 was found 
(Impey et al.\ 1998) using the NICMOS camera, which is supposed to represent 
emission from the quasar host galaxy.

In this letter, we report on the results of near-infrared observations of 
PG~1115+080 made with the 8.2 m Subaru telescope, which was installed 
on the summit of Mauna Kea, Hawaii, at the end of 1998.  Even in the 
very first stage of observations in early 1999 January, nearly 
ultimate performance as a ground-based, large mirror telescope has 
been demonstrated, as discussed in the following. 

Based on $J$ and $K'$ band images taken under an $\sim 0.\hspace{-2pt}''3$ 
seeing condition, we present the detection of the Einstein ring in 
both bands, the distortion of the A1/A2 profiles in the deconvolved 
image, and the structure of the quasar host galaxy.\\

\section
{Observations and Data Reduction}

Photometric observations of PG~1115+080 were made on 1999 Jan 11 and 
13 using the near-infrared camera CISCO (Motohara et al.\ 1998) 
mounted on the Cassegrain port of the 8.2 m 
Subaru Telescope on Mauna Kea.  The camera uses a 1024$\times$1024 
element HgCdTe array (HAWAII; Kozlowski et al.\ 1994; Hodapp et al.\ 1996) with a pixel scale 
of $0.\hspace{-2pt}''116$ pixel$^{-1}$, giving a total field of view of $2'\times 2'$. 
Although the exposure time for the normal full-frame readout is usually no shorter than 
1.5 s/frame, in the present observations we utilized a 
fast-readout mode, by which subarray regions ($64\times 64$ or 
$128 \times 128$ pixels in the corner of each quadrant) could be 
sampled faster, with the minimum being 0.025 s/frame. 

Fortunately, there is, by chance, a reference star (star~A in 
figure~1) located in one of the $128 \times 128$ subarray 
regions while aiming at PG~1115+080 in the other quadrant, so that both 
objects could be observed simultaneously. 


For the cancelation of sky emission we periodically moved the telescope 
in the north--south direction, by $6''$ after taking 96 frames 
at one position with an exposure time of 0.3 s/frame. 
The total numbers of frames obtained were 3456 and 4032 in 
the $J$- and $K'$-band, respectively, giving overall integration 
times of 1037 s ($J$) and 1210 s ($K'$). \\

The obtained data were reduced through standard common processes of sky subtraction, 
flat-fielding, and the correction of bad pixels. After each pixel was divided 
into 4$\times$4 sub-pixels, the images were shifted and added 
by reference to the center of gravity of the stellar image of star~A. Since we 
noticed that the point-spread function (PSF) profiles are not exactly 
the same between the object position and the position of star~A, possibly 
due to the nature of the optical system of CISCO, the PSF of star~A 
was linearly transformed using the matrix with values of elements in table 1,\\

\begin{equation}
\left(
\begin{array}{ll}
cos\ \theta&-sin\ \theta\\
sin\ \theta&cos\ \theta\\
\end{array}
\right)
\left(
\begin{array}{ll}
a&0\\
0&b\\
\end{array}
\right)
\left(
\begin{array}{ll}
cos\ \theta&sin\ \theta\\
-sin\ \theta&cos\ \theta\\
\end{array}
\right)
\end{equation}


\noindent
so as to make component C of PG~1115+080 to be a virtual point source. 
Finally, PSF subtraction and MEM deconvolution were applied to 
the combined image using this transformed PSF. \\

\section
{Results}

Analyses of the PSF of the recorded images in the 
$J$- and $K'$-bands were performed using the IRAF image processing package, 
providing histograms of the statistical distribution of the FWHM values, as 
presented in figure~2. 
The images with FWHMs larger than 0$.\hspace{-2pt}''$45 ($\sim$ 20\% out of the frames 
recorded) were excluded in the shift and add procedure so as to secure 
better images, which are displayed in figures~3a and b. 
The resultant co-added images of star~A are represented by FWHMs of $0.\hspace{-2pt}''36$ 
and $0.\hspace{-2pt}''32$ in the $J$- and $K'$-bands, respectively 
(figures~4a and b). The photometry of each component
was carried out within a circular aperture of 0$.\hspace{-2pt}''$86 diameter. 
In reducing calibrated fluxes of individual sources, flux corrections 
were applied by adding the relative contribution in the outskirt region 
of the transformed PSF, which would, otherwise, be excluded by the 0$.\hspace{-2pt}''$86 
aperture photometry. The result is presented in table~2.


As for the spatial structure of the lensing galaxy, it is assumed that the 
intrinsic radial distribution of the surface brightness has a 
de Vaucouleurs profile (figures~4c and d), and that the 
parameters describing the galaxy have been determined. The actual 
method is, first, to convolve the de Vaucouleurs expression with appropriate 
parameters using the transformed PSF, and then to compare it with the observed 
profile. The results are an effective radius $R_e$ of 
$0.\hspace{-2pt}''58 \pm 0.\hspace{-2pt}''05$, the surface brightness in the center, 
$\mu _e=19.74/17.90 \pm 0.05$ mag arcsec$^{-2}$ (in the $J/K'$ band), 
the ellipticity, $\varepsilon\approx$ 0.1, and a position angle $\theta$ 
of $\sim 65^{\circ}$. 


The photometric error of the zero point is estimated to be $\sim$0.05 mag, 
which is mostly attributed to the uncertainty in the photometry of star~A, whose 
magnitudes were determined by measuring faint standard stars, 
such as FS\ 19 and FS\ 21 of the UKIRT FS Catalog (Casali, Hawarden 1992), with identical
observing configurations. The $J-K'$ colors of the respective components in 
the image are also given in table~2.

It should be noted that component A1 is significantly bluer than 
other split-image components, which may be associated with the fact 
that the fairly large flux ratio in the A1/A2 pair can not be interpreted 
using a normal simple lensing configuration.

A somewhat asymmetric annulus of the Einstein ring is shown to remain when 
subtracting all of the lensed components A1,A2,B, and C and the lensing galaxy G
in both the $J$ and $K'$ bands. 
In presenting the surface brightness of the ring, we select two particular 
regions (R1 and R2 in figure~3e and f) where the brightness of the
Einstein ring suffers less from the uncertainty caused by the process 
of removing the images of the A1/A2 system. 
The results are given in the lower lines of table~2. A two-color image 
produced from figures~3e and f is shown in figure~5a.
The implication of this reddish color is discussed later. 

With the MEM deconvolution task in the IRAF STSDAS package, we deconvolved 
images of the two bands so as to produce a high spatial resolution color image of the quasar,
as presented in figure~5b, where the source size is $0.\hspace{-2pt}''14$ 
in FWHM.  It is important to note that the B and C components are simply 
round shaped, while the A1/A2 system is significantly different from a 
symmetric shape. The meaning of this elongated morphology is briefly interpreted 
in the following discussion. \\


\section
{Discussion}

The spectral energy distributions (SEDs) of the components of the PG~1115+080, 
including the Einstein ring, are plotted in figure~6, where the data 
presented in past literature are added (see inset captions). 
The quasar spectrum appears to be flat (the index of power being nearly zero) 
over the observed wavelength range, except for the blue bump at around the 250 nm 
region. The flat nature of SEDs with a distinctive blue bump is typical 
of quasars with a pole-on geometry (Baker, Hunstead 1995; Baker 1997). The absorption-free 
spectrum indicates that the central source is obscured neither by the host 
galaxy nor by the intervening lensing galaxy. 


On the contrary, the ring regions show considerable red colors, even after 
allowing for the uncertainties, as indicated by the error bars in the 
figure. In the meantime, from a model calculation of the gravitational lens 
system, which is basically the same method described as that by Keeton et al. (1998), 
it is seen that the regions designated by R1 and R2 lie at galactocentric 
distances of $\sim$2~kpc and $\sim$1~kpc, respectively. By comparing the SEDs 
of synthetic model galaxies based on those of nearby galaxies of various types, we can see 
that two-band photometry data of the host galaxy appear to correspond to 
that of an aged galaxy older than 3~Gyr, or even to E/S0 galaxies. Here, we 
refer to the SEDs presented by Coleman et al. (1980) for nearby galaxies; the 
modeled SED was calculated using the code given by Fioc and Rocca-Volmerange (1997), 
in which we adopted an exponential burst star-formation model with a 
time scale of $\tau_{0}$ = 1~Gyr.

It is interesting to note that an inferred age of 3~Gyr corresponds to the 
age of the universe at the quasar redshift. Here, we take $H_0$ = 75 km s$^{-1}$Mpc$^{-1}$ and 
$q_0$ = 0.1 for cosmological parameters. 

The present result deduced from the unexpectedly red color of the host galaxy 
could be interpreted either by 1) a hypothesis that the host galaxy contains 
a certain amount of dust, thus making the color appreciably redder than nearby 
starburst galaxies, or 2) the case that the color is intrinsically red with 
negligible internal absorption. 

In the latter case, however, an age of $>$3~Gyr of the galaxy contradicts an
epoch of $z$ = 1.72 of the Universe. Virtually instantaneous star 
formation with a particular initial mass function (IMF) could ease the 
situation, because the contribution of asymptotic giant branch (AGB) stars 
in the mass range from 5 to 10 $\MO$ makes an SED fairly red on a time 
scale of $\sim$1~Gyr. As for the former case, the dusty environment in the 
host galaxy is not necessarily unrealistic, although the quasar light 
apparently suffers from no significant dust extinction, because the dust in the line of sight could 
be blown off by the intermittent jet activities reported by Michalitsianos et al. (1996).
To clarify the nature of the overall spectrum of the host galaxy, further detailed 
spectroscopic observations are required. 

The observed fluxes at $J$ and $K'$ of the lensing galaxy are plotted in accord 
with a typical SED of the E/S0 galaxy (Coleman et al.\ 1980; Yoshii \& Takahara 1988). It has also been confirmed 
that the obtained radial distribution of the surface brightness is well represented 
by the de Vaucouleurs profile and, moreover, that the shape is slightly, but 
significantly, ellipsoidal, as found by Impey et al. (1998). The fitted position angle 
of the ellipsoid is $\theta \approx 65^{\circ}$, consistent with the result 
of Keeton et al. (1998) in which they report that the position angles of the optical 
axis of ellipsoids generally agree with those of the modeled mass distribution 
in most cases of gravitational-lens systems. In particular, they predicted a 
position angle of PG~1115+0080 as being 67$^{\circ}$, which is the same as the light 
distribution given by the present observation. 

Finally, let us remark that the image profiles of the quasar, exhibit a 
deformed shape of the A1 and A2 components (see figure~5). 
These elongated emission components, whose color is quite similar to that of 
the quasar, itself, is believed to be real in view of the observational accuracy. 
In a quantitative analysis using the model calculation of the present 
lens system, we found that the circumnuclear emission as close as 
$\sim$ 40~pc ($\sim$ 5~mas) to the central quasar should be responsible for 
the bridge-like structure connecting components A1 and A2.
Because the radial distance is considerably larger than the size of the typical broad 
line region ($<$1~pc), we speculate that the light originates 
from illumination caused by electron scattering, supposedly taking place 
near to the surface of the surrounding toric gaseous cloud, from where electrons  
are supplied by intense ionizing radiation of the central source. 
In this view, the exact line of sight to the quasar is presumed to be 
virtually free from absorbing or scattering material; such a configuration is 
consistent with the observed quasar spectrum, as formerly pointed out. \\

\vspace{1pc}\par
The present result, accomplished as one of the first-light observing programs 
of the Subaru Telescope, is indebted to all members of Subaru Observatory, 
NAOJ, Japan.  We would like to express thanks to the engineering staff of Mitsubishi 
Electric Co. for the fine operation of the telescope, and the staff of Fujitsu Co. 
for timely provision of control software. The authors are grateful to M. Fioc and 
B. Rocca-Volmerange for the generous offer to provide their galaxy-modeling 
code, PEGASE.\\

\section*{References}

\re Baker J.C.\ 1997, MNRAS\ 286, 23
\re Baker J.C., Hunstead R.W.\ 1995, ApJ\ 452, L95
\re Barkana R.\ 1997, ApJ\ 489, 21
\re Casali M., Hawarden T.\ 1992, JCMT-UKIRT Newsletter No.4, 33
\re Coleman G.D., Wu C.-C., Weedman D.W.\ 1980, ApJS\ 43, 393
\re Courbin F., Magain P., Keeton C.R., Kochanek C.S., Vanderriest C., Jaunsen A.O., Hjorth J.\ 1997, A\&A\ 324, L1
\re Fioc M., Rocca-Volmerange B.\ 1997, A\&A\ 326, 950
\re Hodapp K.-W., Hora J.L., Hall D.N.B., Cowie L.L., Metzger M., Irwin E., Vural K., Kozlowski L.J. et al.\ 1996, New Astronomy 1, 177
\re Impey C.D., Falco E.E., Kochanek C.S., Leh\'{a}r J., Mcleod B.A., Rix H.-W., Peng C.Y., Keeton C.R.\ 1998, ApJ\ 509, 551
\re Keeton C.R., Kochanek C.S.\ 1997, ApJ\ 487, 42
\re Keeton C.R., Kochanek C.S., Falco E.E.\ 1998, ApJ\ 509, 561
\re Kozlowski L.J., Vural K., Cabelli S.C., Chen C.Y., Cooper D.E., Bostrup G.L., Stephenson D.M., McLevige W.L. et al.\ 1994, Proc.\ SPIE 2268, 353
\re Kristian J., Groth E.J., Shaya E.J., Schneider D.P., Holtzman J.A., Baum W.A., Campbell B., Code A. et al.\ 1993, AJ\ 106, 1330
\re Kundic T., Cohen J.G., Blandford R.D., Lubin L.M.\ 1997, AJ\ 114, 507
\re Michalitsianos A.G., Oliversen R.J., Nichols J.\ 1996, ApJ\ 461, 593
\re Motohara K., Maihara T., Iwamuro F., Oya S., Imanishi M., Terada H., Goto M., Iwai J. et al.\ 1998, Proc.\ SPIE 3354, 659
\re Saha P., Williams L.L.R.\ 1997, MNRAS\ 292, 148
\re Schechter P.L., Bailyn C.D., Barr R., Barvainis R., Becker C.M., Bernstein G.M., Blakeslee J.P., Bus S.J. et al.\ 1997, ApJ\ 475, L85
\re Tonry J.L.\ 1998, AJ\ 115, 1
\re Weymann R.J., Latham D., Angel J.R.P., Green R.F., Liebert J.W., Turnshek D.A., Turnshek D.E., Tyson J.A.\ 1980, Nature\ 285, 641
\re Yoshii Y., Takahara F.\ 1988, ApJ\ 326, 1

\begin{table*}[t]
\begin{center}
Table~1.\hspace{4pt}{Values of elements in matrices}\\
\end{center}
\vspace{6pt}
\begin{tabular*}{\textwidth}{@{\hspace{\tabcolsep}
\extracolsep{\fill}}p{6pc}llll}
\hline\hline\\ [-6pt]
Band&a&b&$\theta$\\
[4pt]\hline\\[-6pt]
$J$ &1.15&1.00&21\\
$K'$&1.12&1.08&33\\
[4pt]\hline\\[-6pt]
\end{tabular*}
\end{table*}

\begin{table*}[t]
\begin{center}
Table~2.\hspace{4pt}{Photometric Results}\\
\end{center}
\vspace{6pt}
\begin{tabular*}{\textwidth}{@{\hspace{\tabcolsep}
\extracolsep{\fill}}p{6pc}cccc}
\hline\hline\\ [-6pt]
Image& \multicolumn{2}{c}{Magnitude$^\ast$}& Color\\
[4pt]\cline{2-3}\\[-6pt]
               & $J$ & $K'$ & $J-K'$\\[4pt]\hline\\[-6pt]
A1 &16.13(16.44) &15.21(15.48) &0.92\\
A2 &16.70(17.01) &15.64(15.91) &1.06\\
B  &18.10(18.41) &17.03(17.30) &1.07\\
C  &17.66(17.98) &16.58(16.85) &1.08\\
G  &17.74(18.75) &15.98(16.88) &1.76\\
R1 &21.74$\pm$0.28$^\dagger$        &19.23$\pm$0.06$^\dagger$ &2.51\\
R2 &21.31$\pm$0.18$^\dagger$        &19.38$\pm$0.06$^\dagger$ &1.93\\[4pt]\hline
\end{tabular*}
\vspace{6pt}\par\noindent
$*$ Corrected total magnitude (see text). 
The magnitudes inside parentheses are measured in a $0.\hspace{-2pt}''86$ 
diameter aperture.
\par\noindent
$\dagger$ Surface brightness (mag arcsec$^{-2}$).
\end{table*}

\onecolumn
\begin{figure}[p]
\epsfxsize=17cm
\caption{Full-frame image of PG~1115+080 and the reference 
star (star~A). The small squares correspond to subarray sections 
in the fast readout observing mode. The object and star~A were 
observed simultaneously.}
\end{figure}
\begin{figure}[p]
\epsfxsize=17cm
\epsfbox{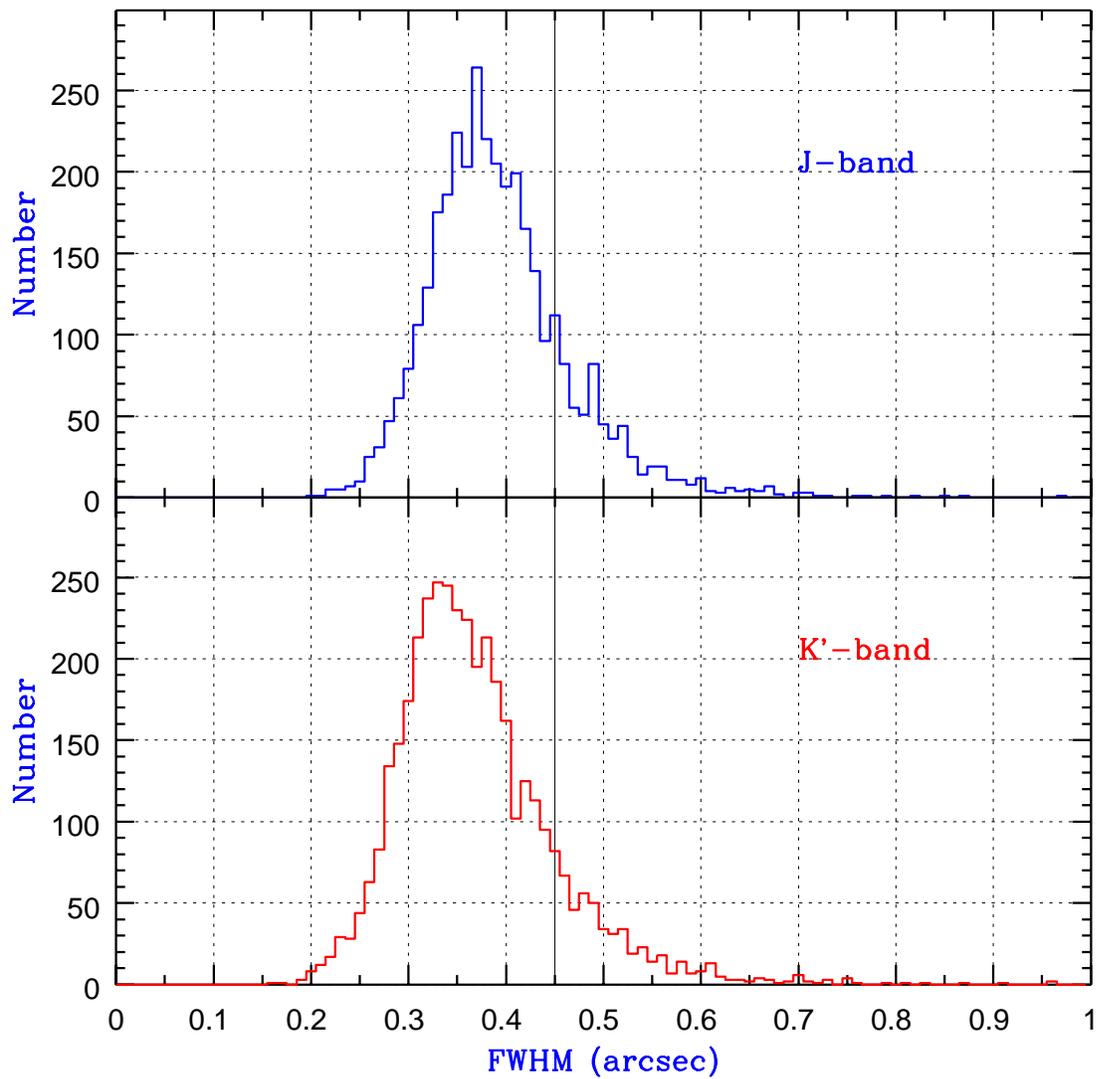}
\caption{Distribution of the FWHM of star~A during the 
observation.  Only images whose FWHM was smaller 
than 0$.\hspace{-2pt}''$45 were used to make the combined image.}
\end{figure}
\begin{figure}[p]
\epsfxsize=6.5cm
\hspace*{2cm}\epsfbox{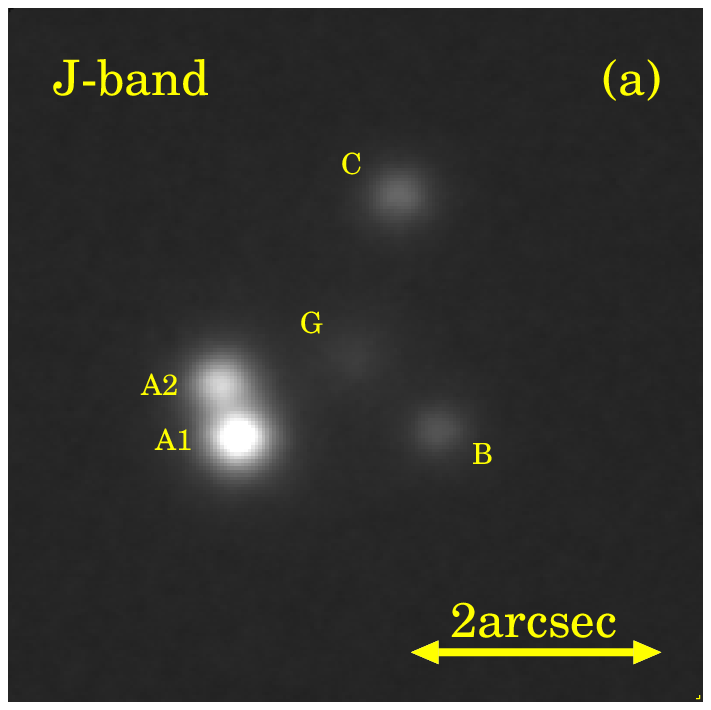}
\epsfxsize=6.5cm
\hspace*{2mm}\epsfbox{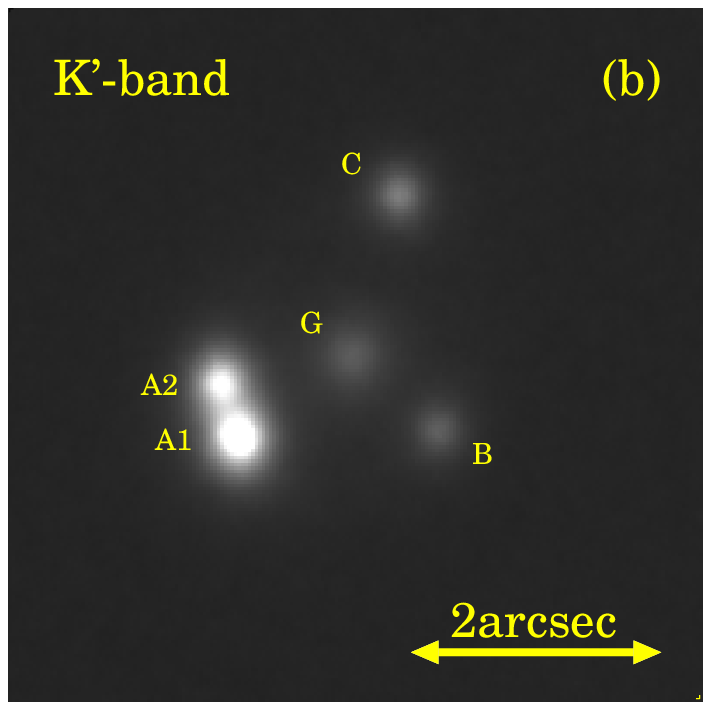}

\epsfxsize=6.5cm
\hspace*{2cm}\epsfbox{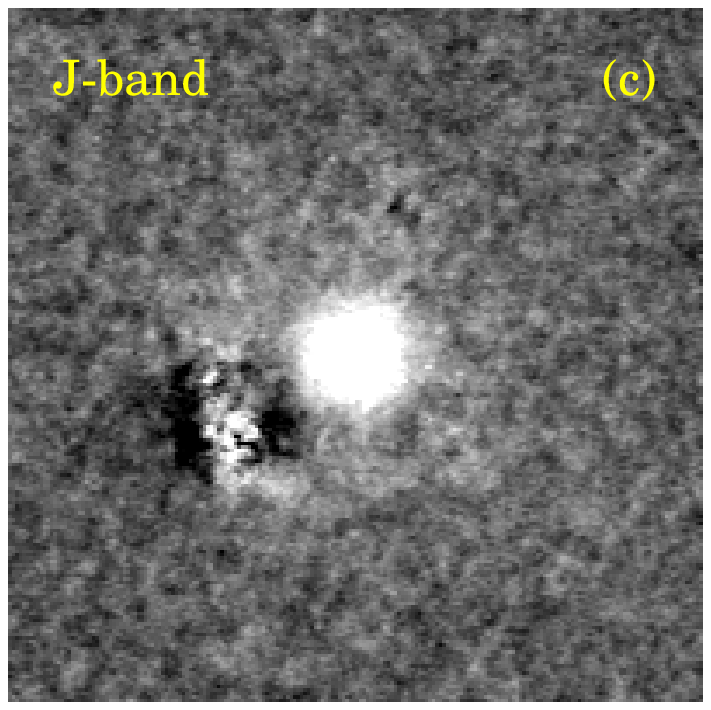}
\epsfxsize=6.5cm
\hspace*{2mm}\epsfbox{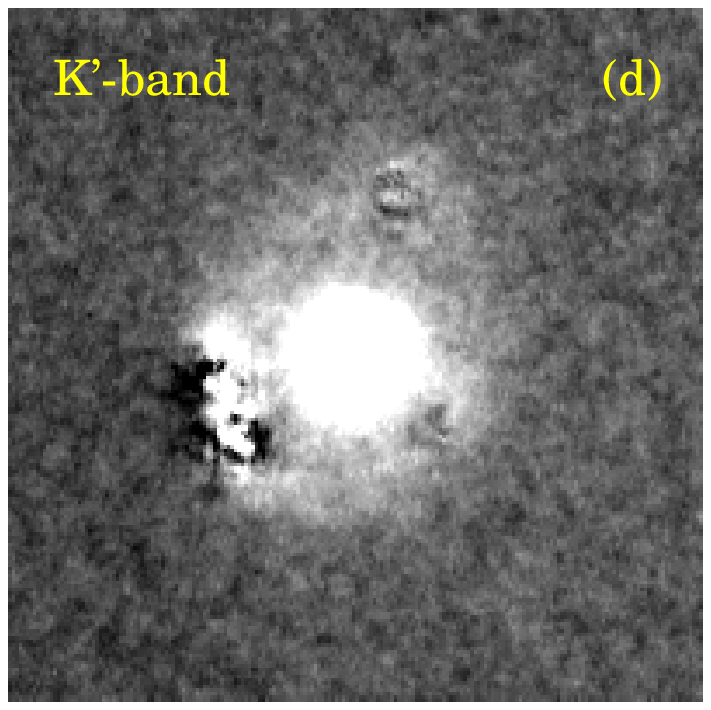}

\epsfxsize=6.5cm
\hspace*{2cm}\epsfbox{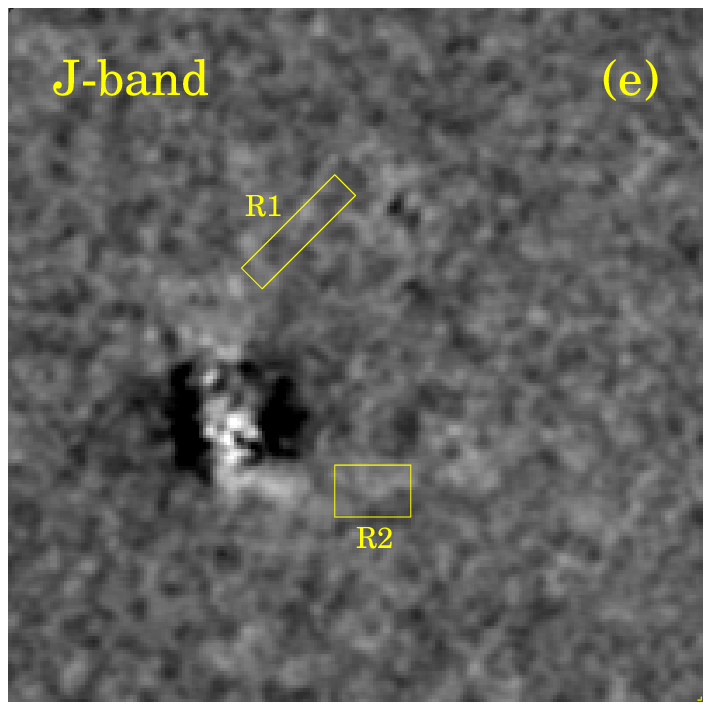}
\epsfxsize=6.5cm
\hspace*{2mm}\epsfbox{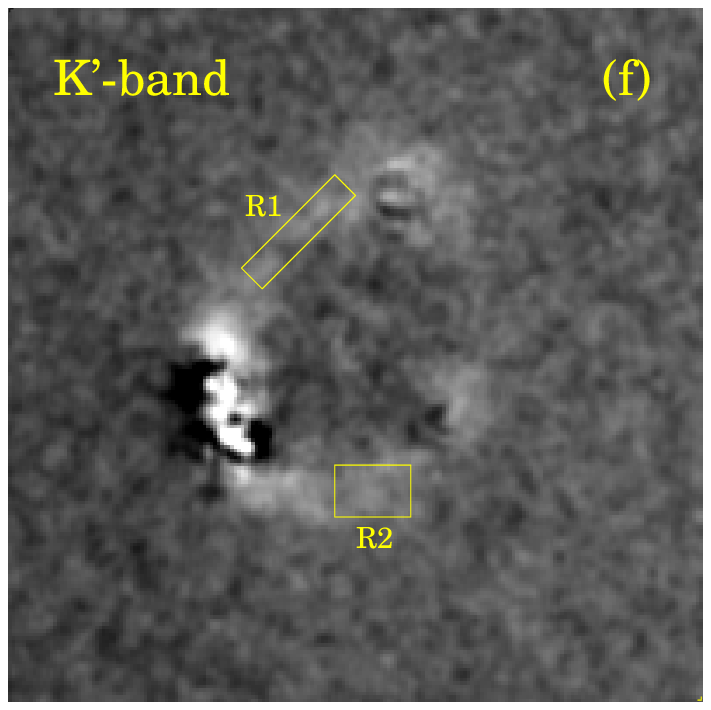}
\caption{Combined images and the PSF subtraction processes. 
The ratio of the display range between the corresponding $J$ 
and $K'$-band image is fixed to the flux ratio of the quasar in these 
bands. (a)-(b) The combined images in the $J$ and $K'$-bands.
(c)-(d) The image of the lensing galaxy after subtraction of the quasar 
component by using the transformed PSF. (e)-(f) The Einstein
ring visible in the frame when the lensing galaxy image obtained 
by the convolved de Vaucouleurs profile (figure~4) is removed. 
R1 and R2 indicate the region where the surface brightness was measured.}
\end{figure}
\begin{figure}[p]
\epsfxsize=17cm
\epsfbox{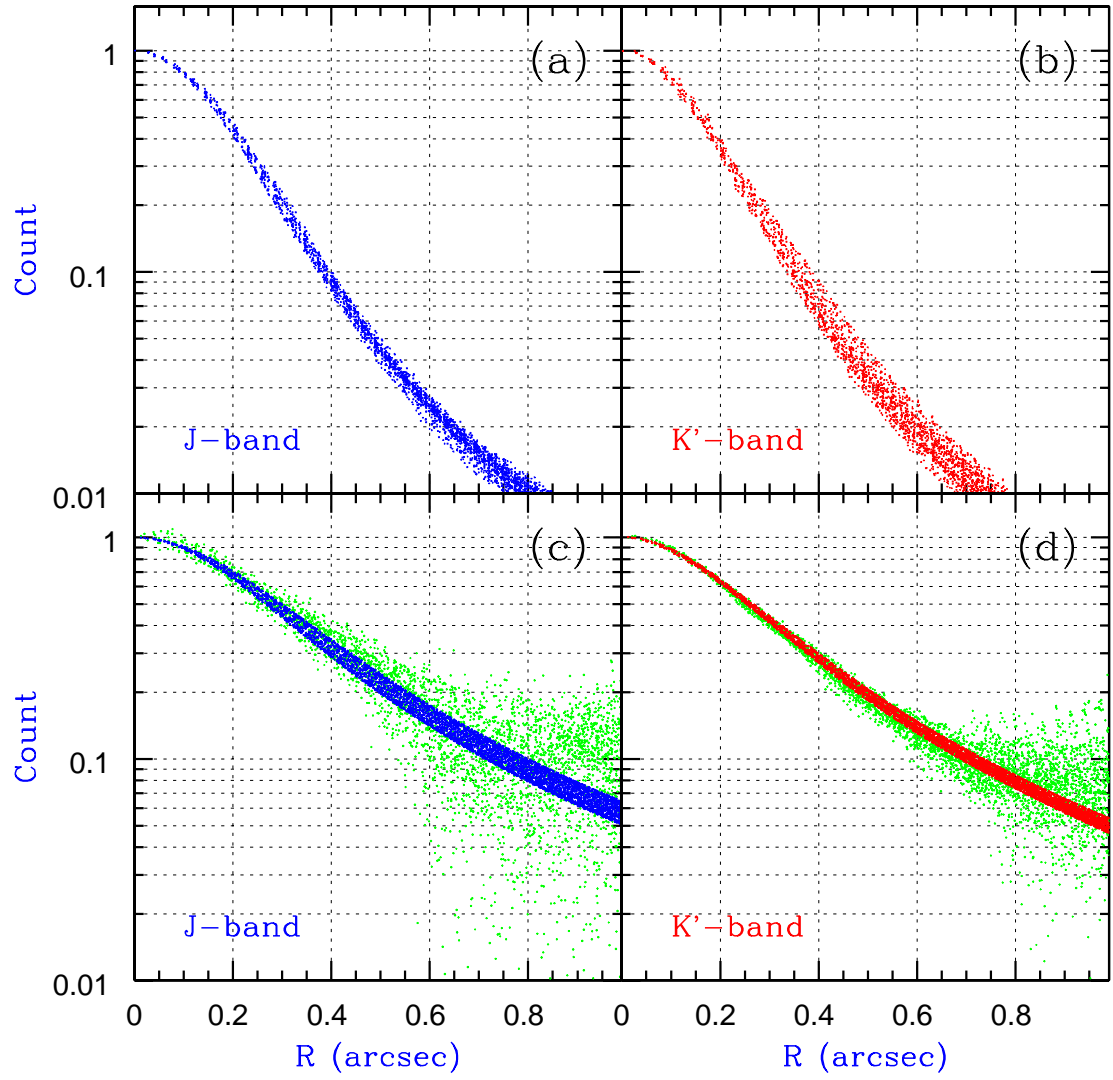}
\caption{(a)-(b) Radial profile of the combined image of star~A. 
The FWHMs are $0.\hspace{-2pt}''36$ ($J$) and $0.\hspace{-2pt}''32$ ($K'$). 
(c)-(d) Radial profile of the lensing galaxy and the fitted model of the 
PSF convolved de Vaucouleurs profile (see text).}
\end{figure}
\begin{figure}[p]
\epsfxsize=10cm
\hspace*{4cm}\epsfbox{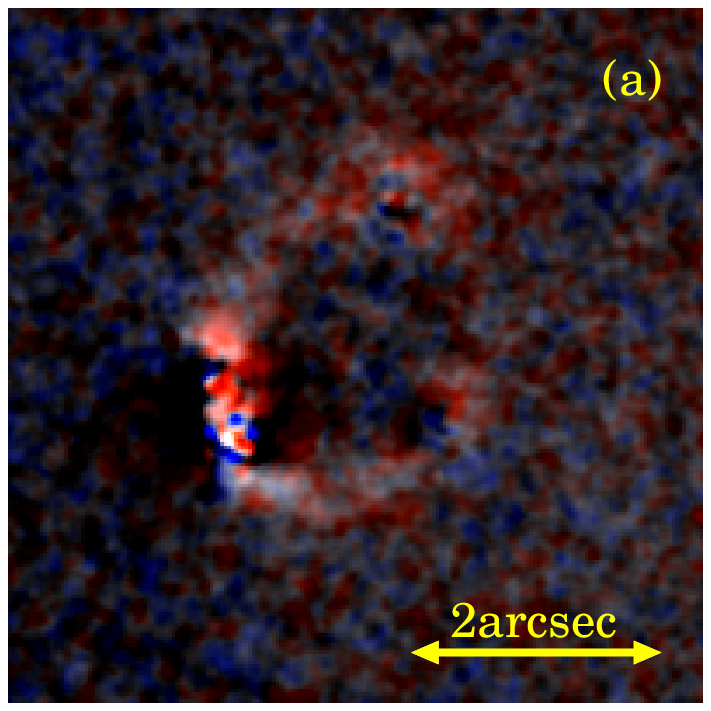}

\epsfxsize=10cm
\hspace*{4cm}\epsfbox{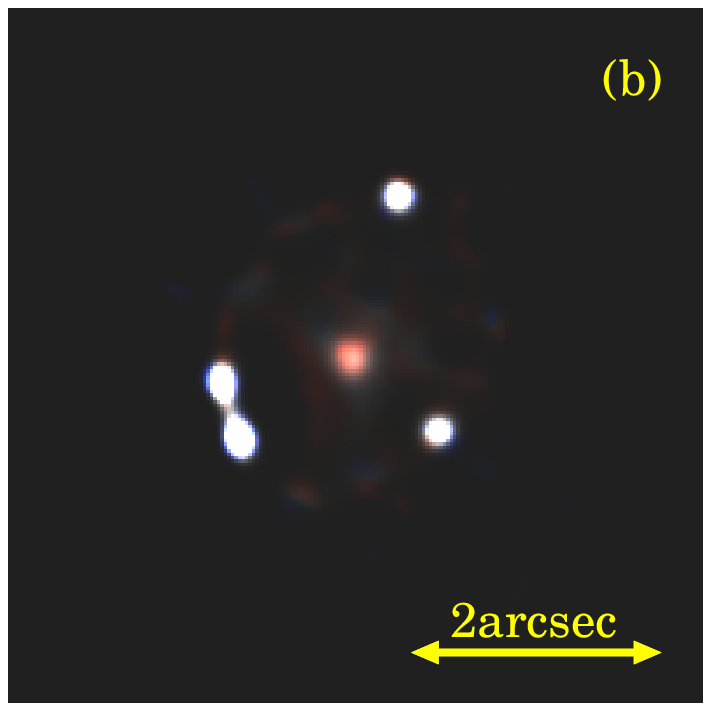}
\caption{$J$, $K'$ 2 color image of (a) the Einstein ring and 
(b) the deconvolved image (FWHM $0.\hspace{-2pt}''14$). Both images are 
displayed to make the color of component C white.}
\end{figure}
\begin{figure}[p]
\epsfxsize=10cm
\hspace*{4cm}\epsfbox{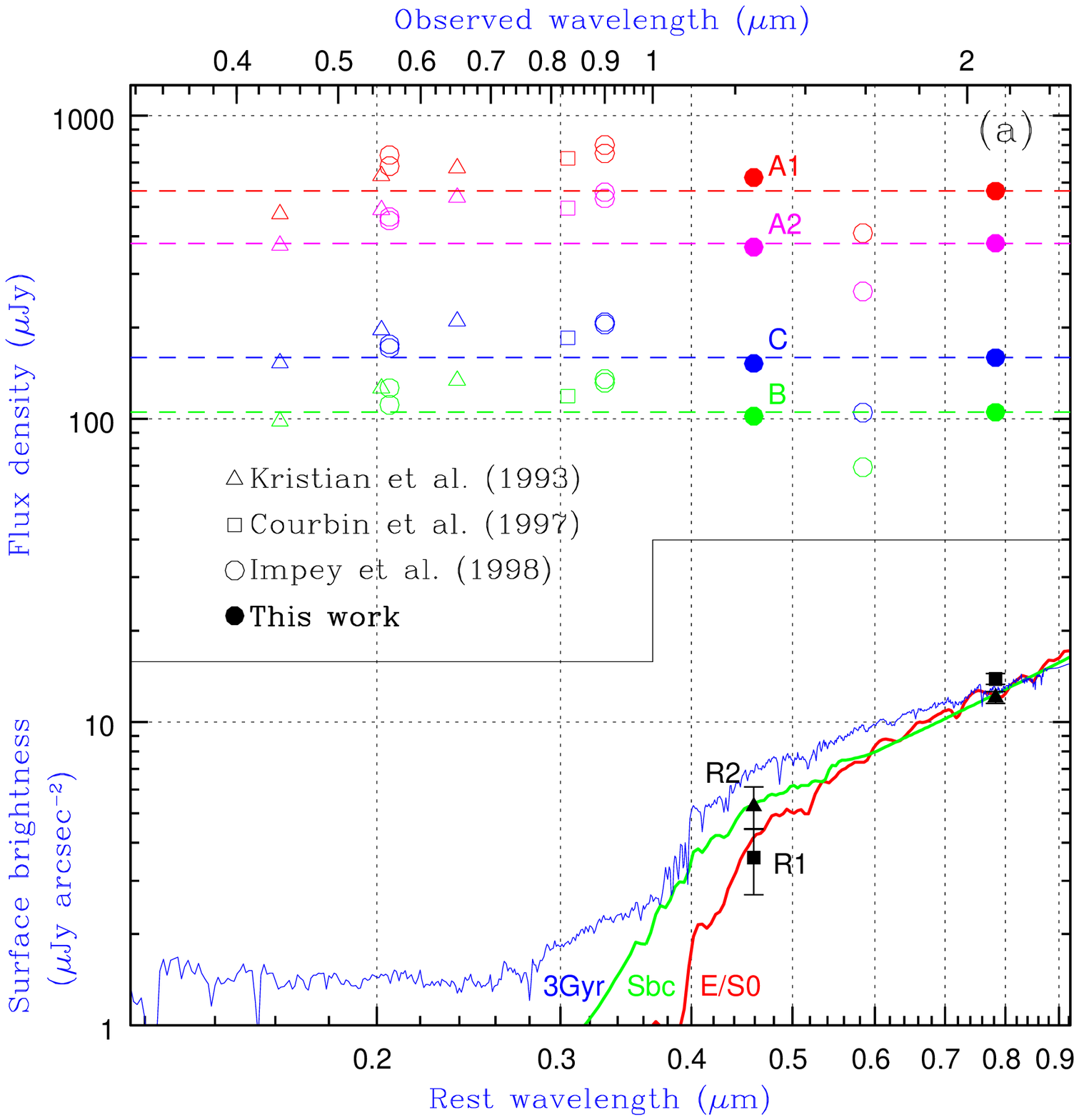}

\epsfxsize=10cm
\hspace*{4cm}\epsfbox{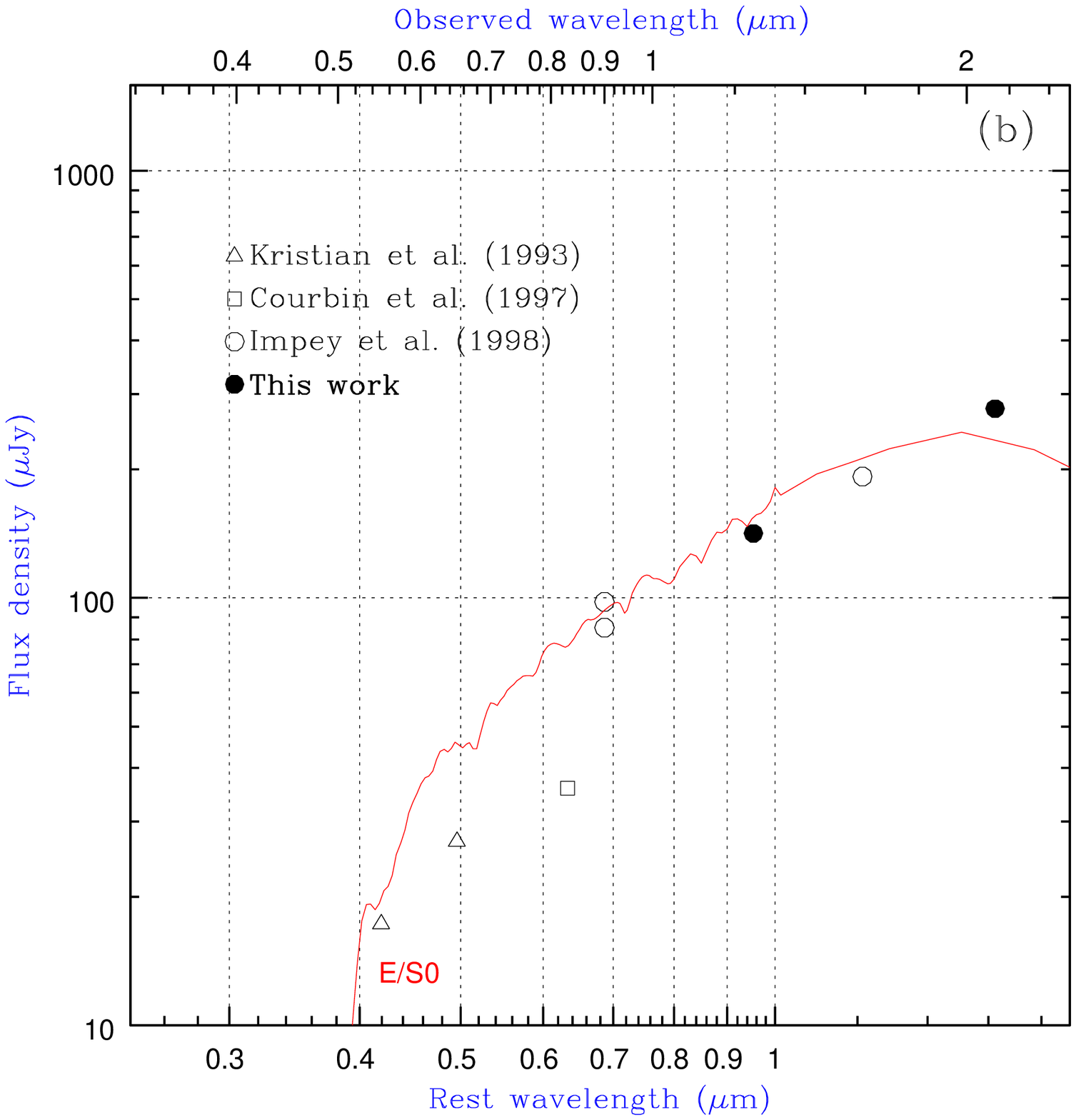}
\caption
{(a) SEDs of each lensed component and the Einstein ring. 
The thick solid lines represent the typical SEDs for E/S0 and Sbc galaxies 
(Coleman et al. 1980) and the thin solid line indicates the SED of the 
synthetic model galaxy (Fioc, Rocca-Volmerange 1997) with an age of 
3 Gyr. (b) The SED of the lensing galaxy with the typical SED for 
E/S0 (Coleman et al. 1980; Yoshii, Takahara 1988).}
\end{figure}

\end{document}